\documentclass[]{svjour2}                    

\usepackage{graphicx}
\usepackage{amssymb}
\usepackage{amsfonts}
\usepackage{amsmath}
\usepackage{setspace}

\renewcommand{\d}{\mathrm{d}}
\newcommand{\e}{\mathrm{e}}

\journalname{Journal of Statistical Physics}
\begin{document}

\title{Noise filtering strategies of adaptive signaling networks}
\subtitle{Application to {\it E. coli} chemotaxis}

\author{Pablo Sartori \and Yuhai Tu}

\institute{P. Sartori \at
              Max Planck Institute of Complex Systems \\
              \email{pablo.sartori@gmail.com}
           \and
           Y. Tu \at
              IBM T.J. Watson Research Center \\
              Tel.: +123-45-678910\\
              \email{yuhai@us.ibm.com}
}

\date{Received: 1st December, 2010 / Accepted: 1st December, 2010}

\maketitle

\begin{abstract}

Two distinct mechanisms for filtering noise in an input signal are identified in a class of adaptive sensory networks. We find that the high frequency noise is filtered by the output degradation process through time-averaging; while the low frequency noise is damped by adaptation through negative feedback. Both filtering processes themselves introduce intrinsic noises, which are found to be unfiltered and can thus amount to a significant internal noise floor even without signaling. 
These results are applied to \textit{E. coli} chemotaxis. We show unambiguously that the molecular mechanism for the Berg-Purcell time-averaging scheme is the dephosphorylation of the response regulator CheY-P, not the receptor adaptation process as previously suggested. The high frequency noise due to the stochastic ligand binding-unbinding events and the random ligand molecule diffusion is averaged by the CheY-P dephosphorylation process to a negligible level in {\it E.coli}.
We identify a previously unstudied noise source caused by the random motion of the cell in a ligand gradient. We show that this random walk induced signal noise has a divergent low frequency component, which is only rendered finite by the receptor adaptation process. For gradients within the {\it E. coli} sensing range, this dominant external noise can be comparable to the significant intrinsic noise in the system. The dependence of the response and its fluctuations on the key time scales of the system are studied systematically.  We show that the chemotaxis pathway may have evolved to optimize gradient sensing, strong response, and noise control in different time scales.

\end{abstract}

\section{Introduction and summary}
\label{intro}

A main function of biological sensory systems is to detect minute signals in fluctuating environments. One key task of the underlying signaling pathways (networks) is then to mitigate the effects of external noise. However, the network itself can introduce noise 
due to the finite number of bio-molecules involved in the intracellular signaling process. 
Both the intrinsic stochasticity of the signaling networks and the noise in the input can contribute to large fluctuations in the output of the system. Therefore, correctly identifying the dominant sources of noise and deciphering the cellular strategy in noise filtering 
are critical in understanding the proper functioning of biological sensory systems.

In their now classical work \cite{berg}, Berg \& Purcell proposed a time-averaging scheme for filtering noise from the stochastic ligand receptor binding process. The key idea was that if the noise correlation time is $\tau_s$, the output variance can be reduced by averaging the signal over a longer timescale $\tau_{\langle s\rangle}$ according to $\sigma^2 \propto \tau_s/\tau_{\langle s\rangle}$. There has been much recent progress in rigorous derivations of the correlation time $\tau_s$ of the diffusion and binding noise for different systems \cite{bialek,nwgprl,nwgprlramp,levine,nwgeuk}. However, the Berg-Purcell expression itself has been given for granted, and the molecular origin of the time averaging mechanism remains unclear. No connection has been established between  $\tau_{\langle s\rangle}$, the dynamics of the pathway and its many time scales. Furthermore, often an \textit{ad hoc} value of $\tau_{\langle s\rangle}$, usually the longest time scale of the system, is used without justification  \cite{berg,bialek}. One goal of this paper is to understand and clarify the molecular origin of the Berg-Purcell time-averaging mechanism based on rigorous analysis of the stochastic dynamics for a class of adaptive sensory systems. 

Most sensory signaling systems, such as bacterial chemotaxis, adapt \cite{bergadap,sensadap,voscience} through feedback control \cite{yietal} to maintain a high sensitivity over a wide range of backgrounds \cite{mellotuwr}. A step stimulus (input) triggers a fast change of the output (response) followed by a slow recovery (adaptation) to its pre-stimulus level. Given adaptation's long timescale and its stabilizing effect, a natural question is whether it also serves as a noise filter. More specifically, whether the adaptation time serves as the averaging time $\tau_{\langle s\rangle}$. This is indeed assumed to be the case by several previous studies, particularly for {\it E. coli} chemotaxis \cite{nwgprlramp,piglesias}. Here, we show unambiguously that the Berg-Purcell averaging time is {\it not} the adaptation time. In fact, $\tau_{\langle s\rangle}$ is the response time that is controlled by the signal degradation process. This is the first result of this paper. 

The time-averaging mechanism works well in filtering high frequency noise with correlation time shorter than the response time. Indeed, for {\it E. coli} chemotaxis, the noises due to the stochastic ligand receptor binding-unbinding process and the random ligand diffusion, considered by Berg \& Purcell originally \cite{berg} and followed by other more recent studies \cite{berg,bialek,nwgprl,nwgprlramp,levine,nwgeuk}, have short time scales ($\sim10^{-5}$s for ligand diffusion and $\sim10^{-7}$s for ligand binding) and is rendered irrelevant by averaging over a relatively long response time ($\tau_{\langle s\rangle}\sim 0.1$s). For {\it E. coli} chemotaxis the dominant signal fluctuation, we find, comes from the random motion of the cell. In a ligand gradient, this random motion introduces an unbounded low frequency fluctuation in the input signal. Such random walk induced signal noise has not been studied before, even in recent works regarding chemotaxis noise in a ligand gradient \cite{nwgprlramp}. Here, we show that random walk noise is not filtered by the time-averaging mechanism. Instead, this low frequency noise is suppressed by the receptor adaptation process by a mechanism general to adaptive networks which we call feedback control. The frequency (or time) dependence in the filtering function in the case of adaptive control is opposite to that of the time-averaging mechanism. This is the second result of this paper.

However, these intracellular biochemical noise filtering mechanisms, time-averaging through output degradation and feedback control through receptor adaptation, also introduce fluctuations in the output themselves due to the finite number of molecules involved in these processes. Here, we derive the expressions of these intrinsic noises from the stochastic pathway dynamics. We show that both these intrinsic noises are not filtered (suppressed) by the pathway itself and can thus contribute to a significant intrinsic noise floor even in the absence of any signal. This is the third result of this paper.  

Finally, we verify these analytical results by simulating the propagation of signal noise in the nonlinear {\it E. coli} chemotaxis pathway dynamics for bacterial cells moving in an attractant gradient. Using biologically relevant parameters, we show that the random-walk induced noise is the main source of signal noise, while the ligand diffusion and binding noises are irrelevant. We also systematically study the dependence of the response and its fluctuation (due to both extrinsic and intrinsic noises) on the two key time scales (response time and adaptation time). Form our analysis, we argue that the chemotaxis pathway has evolved to optimize gradient sensing, strong response, and noise control in different time scales.

\section{Two noise filtering strategies of adaptive signaling networks}
\label{sec:1}

We study noise in a generic three-node adaptive network (Fig. 1A, see \cite{yuhaiberg,chaotang}) that abstracts the {\it E. coli} chemotaxis pathway (Fig. 1B). We will then show that the results can be further generalized to other related adaptive signaling network topologies shown in Fig. 1C-F.

In Fig. 1A, the signal $s$ is sensed by the fast ``activity'' node $a$, which controls the output $y$. The ``memory'' $m$, which depends on $a$, carries out adaptation by feeding back to control the activity $a$. The noisy dynamics of this adaptive network can be described by the Langevin equations:
\begin{align}
\mathrm{Activity}\;\;\;\;\;\;\frac{\d a}{\d t} &=f_a(a,m,s+\eta_s)+\eta_a;\notag\\
\mathrm{Adaptation}\;\;\;\;\;\frac{\d m}{\d t} &=f_m(a,m)+\eta_m;\notag\\
\mathrm{Output}\;\;\;\;\;\;\frac{\d y}{\d t} &=f_y(a,y)+\eta_y.
\label{interact}
\end{align}
where $\eta$ represents different noise sources: $\eta_a$, $\eta_m$ and $\eta_y$ are the internal white noises caused by the stochastic biochemical reactions; the spectrum of the signal noise $\eta_s$ depends on its origin as described later.

The interactions among the nodes are characterized in Eq. (\ref{interact}) by the functions $f$, whose exact forms  can be obtained from the detailed kinetics of the system. The network topology (Fig. 1A) constrains these interactions in the following way. The feedback mechanism requires $f_{a,m}f_{m,a}<0$; the stability of the system requires $f_{a,a}\leq0$, $f_{y,y}\leq0$ and $f_{m,m}\leq0$; and accurate adaptation is achieved by choosing $f_{m,m}=0$ \cite{mellotupa}. With these constrains, Eq. \ref{interact} generally defines an adaptive system with response rate $\omega_y=|f_{y,y}|$ and adaptation rate $\omega_m=|f_{a,m}f_{m,a}/f_{a,a}|$. Here $_{,x}$ stands for derivative with respect to $x$. Also, for each characteristic frequency $\omega_x$ we define a characteristic time as $\tau_x\equiv1/\omega_x$.

There are two types of signal related noise. The first is caused by the stochasticity of the discrete signal sensing events such as ligand binding-unbinding, which was considered originally by Berg and Purcell \cite{berg}. This ligand binding noise exists even for constant signals. The second signal noise is caused by the random temporal variations of the external signal, e. g., air pressure fluctuations for ear hair cells, diffusion of ligand around a chemoreceptor, or fluctuations in attractant concentration sensed by a bacterium due to the cell's random motion up an attractant concentration gradient. This latter case will be studied in detail in the next section. In general, we model the signal noise $\eta_s$ as exponentially correlated with strength $\langle\eta_s^2\rangle$ and correlation time $\tau_s$.
 	
For relatively small noise, Eq. \ref{interact} can be linearized around its steady state and the transfer function for each of the noise sources can be determined analytically in frequency ($\omega$) space. For example, for signal noise one has $\delta \hat{y}(\omega)=\chi(\omega)\hat{\eta}_s(\omega)$, where
\begin{align}
\chi(\omega)=i A_s \omega \omega_y/[(i\omega+\omega_m)(i\omega+\omega_y)]
\end{align}
is the transfer function for small signals, with $A_s$ dependent on the interaction functions \cite{yuhaiberg}. The output $y$ in a single cell fluctuates due to both external signal noise and intrinsic pathway noise. The strength of the output fluctuation can be characterized by its variance, which can be calculated by $\sigma_Y^2=(2\pi)^{-1}\int_{-\infty}^{\infty}\langle\delta\hat{y}(\omega)^2\rangle\d\omega$. Summing up contributions from all the noise sources, we obtain a compact expression of the total variance for the output $y$:
\begin{equation}
\sigma_ Y ^2=C_s\times \frac{\tau_s}{\tau_{\langle s\rangle}}
+C_a\times \frac{\tau_a}{\tau_{\langle a\rangle}}
+C_m\times \frac{\tau_m}{\tau_{\langle m\rangle}}
+C_y\times \frac{\tau_y}{\tau_{\langle y\rangle}},
\label{variance}
\end{equation}
where each contribution, originating from a separate node in the network, has been written in the form of a dimensionless noise amplitude, defined as $C_x=A^2_x\langle \eta_x^2\rangle\tau_x/2$ with $A_x$ the corresponding transfer function amplitude; multiplying the ratio of the characteristic time of the node $\tau_x$ over a node-specific averaging-time $\tau_{\langle x\rangle}$ which depends on the key timescales of the system. The times for the signal noise and the internal noises are
\begin{align}
\tau_{\langle s\rangle}&=(\tau_y+\tau_m)(\tau_y+\tau_s)(\tau_m+\tau_s)/\tau_m^2;\\
\tau_{\langle a\rangle}&=\tau_y(\tau_m+\tau_y)/\tau_m;\notag\\
\tau_{\langle m\rangle}&=\tau_y+\tau_m;\notag\\
\tau_{\langle y\rangle}&=\tau_y.\notag
\end{align}

Eq. (\ref{variance}) presents a comprehensive picture of the different sources of noise in the system and their contributions to the output variance. The forms of the averaging times reveal the underlying mechanisms for filtering different forms of noise. From the explicit expression of the signal averaging time $\tau_{\langle s\rangle}$ (Eq. 4), two distinct noise filtering mechanisms and their responsible underlying biochemical processes are revealed.

\paragraph{1) Time averaging:} For high frequency noise ($\tau_s\ll \tau_y,\tau_m$) one has $\tau_{\langle s\rangle}\approx\tau_y(1+\tau_y/\tau_m)\approx \tau_y$. The last approximation is justified since in most adaptive sensory systems the initial response to a signal is much faster than the adaptation $\tau_y\ll\tau_m$ (examples in \cite{sensadap,voscience} all substantially more than ten fold). Therefore, we have
\begin{align}
\sigma_Y^2\approx C_s \tau_s/\tau_y,
\label{timave}
\end{align}
which follows exactly the Berg-Purcell time-averaging scheme. More importantly, through the rigorous derivation of the time-averaging scheme, its underlying molecular mechanism is revealed. From Eq. \ref{timave}, the averaging time is $\tau_y$, the response time, which is essentially the decay time of the output. Our finding here unambiguously shows that the underlying mechanism for averaging high frequency noise is the output degradation process. The slow adaptation dynamics does not play any role in time-averaging, contrary to what was proposed before \cite{nwgprlramp,piglesias}. For {\it E. Coli}, we will see that output degradation is slow enough as to render irrelevant ligand diffusion and binding noise.

\paragraph{2) Adaptive control:} For low frequency noise ($\tau_s\gg \tau_m,\tau_y$) one has $\tau_{\langle s\rangle}\approx\tau_s^2(1+\tau_y/\tau_m)/\tau_m\approx \tau_s^2/\tau_m$, which is no longer an averaging time. Indeed, the output variance takes the form
\begin{align}
\sigma_Y^2\approx C_s\tau_m/\tau_s,
\label{adapcon}
\end{align}
fundamentally different from the Berg-Purcell scheme. $\sigma_Y^2$ is now inversely proportional to $\tau_s$ and proportional to $\tau_m$: faster adaptation leads to smaller output fluctuation. This noise filtering mechanism, which we call adaptive noise control, is carried out by the adaptation process and it only applies when adaptation is faster than the correlation time of the noise. Later, we will show how this mechanism controls the effects of signal noise caused by the random walk (run-tumble) motion of {\it E. coli} cells.

In addition to the noise introduced by the signal $s$, each internal node of the network ($a$, $m$ and $y$) contributes to the output fluctuation due to the intrinsic stochasticity of the intracellular biochemical reactions. The time-averaging picture (Eq. \ref{variance}) is useful to characterize internal noises. As the readout of the signal $s$ by $a$ is fast, $\tau_{\langle a\rangle}\sim\tau_y\gg\tau_a$, and the activity noise is heavily damped. The noise from the adaptation node $m$ goes as $\sim(1+\tau_y/\tau_m)^{-1}$, which is almost unfiltered since $\tau_m\gg\tau_y$ in sensory systems. The noise from the output node is always unfiltered, as $\tau_{\langle y\rangle}=\tau_y$. These two unfiltered internal noise sources, which exist even in the absence of the external signal, constitute the noise floor of the signaling network. Their absolute and relative strengths depend on the details of the system, and will be discussed later in the case of {\it E. coli} chemotaxis.

So far we have focused on the network of Fig. 1A, which is an abstraction of the {\it E. coli} chemotaxis pathway (Fig. 1B). We now proceed to extend our results to other adaptive topologies (Fig. 1C-F). It has recently been shown that the main three-node network topologies that can exhibit robust perfect adaptation are the one of chemotaxis shown in Fig. 1A, and that in Fig. 1C \cite{chaotang}. The topology in Fig. 1C is named \textit{Incoherent Feedforward Loop with a Proportioner Node} (IFFLP) in \cite{chaotang}. For the chemotaxis topology, the condition for accurate adaptation is simply $f_{m,m}=0$ as shown before, whereas the IFFLP network requires a more stringent condition $f_{y,m}f_{m,a}=f_{y,a}f_{m,m}$. If the perfect adaptation condition is satisfied by the IFFLP network, it is not hard to show that the transfer function has the same form as that of the chemotaxis topology. The only difference is that the adaptation and response rates for the IFFLP network are given by  $\omega_m=|f_{m,m}|$ and $\omega_y=|f_{y,y}|$ respectively. Since the transfer function determines the signal filtering properties of the system, the two noise filtering strategies here described hold exactly for the alternative IFFLP topology.

In general, any adaptive three-node topology that exhibits accurate adaptation will do so by satisfying either the condition of the IFFLP network or that of the chemotaxis network \cite{chaotang}. For example the Yeast osmosensing and the olfactory adaptation topologies (Figs. 1D\&E) have to satisfy the chemotaxis condition ($f_{m,m}=0$) to exhibit accurate adaptation. On the other hand, the topology in Fig. 1F exhibits accurate adaptation if it satisfies the IFFLP condition ($f_{y,m}f_{m,a}=f_{y,a}f_{m,m}$). When any of these topologies satisfies the adequate condition and exhibits perfect adaptation, the transfer function has the form of the chemotaxis transfer function. As a consequence, the two noise filtering strategies found in this paper hold true for any three-node topology that exhibits accurate adaptation.


\section{ {\it E. coli} chemotaxis: noise, filtering, and design trade-offs}
\label{sec:1}

We now apply the general results of the previous section to the case of {\it E. coli} chemotaxis pathway, where the interaction functions and the noise strengths can determined based on the underlying biochemical reactions. In the following we present a simple model for bacterial chemotaxis following the recent work by Tu, Shimizu and Berg \cite{yuhaiberg} before addressing the noise effects in the system.

Over a wide range of ligand attractant concentration $[L]$, the signal an {\it E. coli} cell senses depends logarithmically on $[L]$: $s=\ln([L]/K_I)$, with a characteristic dissociation constant $K_I$ ($K_I\approx 18\mathrm{\mu M}$ for MeAsp as considered in this paper). The kinase activity of the chemoreceptor complex is given by $a$, and $m$ is the methylation level of the chemoreceptor. The output $y$ is the number of CheY-P molecules. A coarse-grained model of the chemotaxis pathway was proposed \cite{yuhaiberg} and verified \cite{shimitu} with interaction functions given by:
\begin{align}
f_a(a,m,s)=-\omega_a\left(a-\frac{1}{1+\mathrm{e}^{E(m,s)}}\right )
\label{fa}
\end{align}
with $\omega_a=50\mathrm{Hz}$, and $E(m,s)=N[\alpha(m- m_0)-s]$ the free energy of a $N$-receptors cluster with $N=6$, $\alpha=2$, and $m_0=1$;
\begin{align}
f_m(a,m)=F(a)
\label{fm}
\end{align}
has a root at $a_0=1/3$ and a negative slope which results in an adaptation time of $\tau_m=10\mathrm{s}$; and
\begin{align}
f_y(a,y)=k\mathcal{N}a(y_T-y)-\omega_z y
\label{fy}
\end{align}
with $\mathcal{N}=600$ independent receptor units with a phosphate transfer rate $k=3\times 10^{-3}\mathrm{Hz}$ from CheA-P to the pool of $y_T=10^4$ CheY molecules, and a CheY-P decay rate assisted by the phosphatase CheZ of $\omega_z=1.3\mathrm{Hz}$ which results in a response time of $\tau_y=0.5\mathrm{s}$.

The noise strengths can be obtained by summing the rates in the master equation underlying the Langevin dynamics \cite{vankampen}. For the activity switching dynamics, the noise is binomial and its strength is given as $\langle\eta_a^2\rangle=2a(1-a)\omega_a/\mathcal{N}$. For the methylation/demethylation processes, $\langle\eta_m^2\rangle=2a(1-a)F'(a)/\mathcal{N}$ because the adaptation dynamics depends only on the activity. Since the fluctuation of the number of CheY-P molecules in a cell is due to an underlying Poisson process, we have $\langle\eta_y^2\rangle=2y\omega_z$.

By using both the general analytical results from last section and direct simulations of the Langevin equation (Eq. (\ref{interact})), we have studied the contributions to CheY-P level fluctuation from different noise sources (external and internal). Different noise contributions and their dependence on the two key time scales of the system $\tau_m$ and $\tau_z$ are summarized in Figure 2.

We first consider the high frequency noise from ligand binding and diffusion, which have been the focus of most previous works \cite{berg,bialek,nwgprl,nwgprlramp}. The signal noise $\eta_d$ caused by the diffusion of ligand molecules around a unit of $N$ receptors of size $l$ can be modeled \cite{bialek,landau2}. It is not hard to show that the ligand diffusion noise $\eta_{d}$ can be characterized by a strength $\langle\eta^2_{d}\rangle\approx4(2\pi)^{5/2}/(\tau_d[L](Nl)^3)$ and a correlation time $\tau_{d}\approx (Nl)^2/D\sim 10^{-5}\mathrm{s}$, where the ligand diffusion constant $D\approx10^{-5}\mathrm{cm}^2/\mathrm{s}$ is used for MeAsp, and $l\approx 0.01\mu$m  is used for receptor Tar \cite{berg,recsize}). For {\it E. coli} chemo-receptors, the signal noise $\eta_{b}$ can also be characterized by its strength $\langle\eta^2_b\rangle\approx (1-a)(K_i+[L])^2/(\tau_bK_i[L])$ from the stochastic binding-unbinding process with an even shorter time scale $\tau_b\approx (DNl(K_i+[L]))^{-1}\sim 10^{-7}\mathrm{s}$. According to our analysis, these high frequency noises are averaged over the response time $\tau_z$, but not the adaptation time $\tau_m$. These general predictions (solid lines in Fig. 2) are confirmed by direct simulations (symbols in Fig. 2) of the pathway dynamics (Eqs. 7-9). As shown in Fig. 2A, the contribution to the output fluctuation from the ligand diffusion noise (light blue symbols and line) and ligand binding noise (purple symbols and line) decreases with $\tau_z$; while they are independent of $\tau_m$ as shown in Fig. 2B. Given that the averaging time is much longer than the ligand binding and diffusion times  $\tau_y=0.5 \mathrm{s}\gg\tau_{d,b}$, the effects of the ligand binding and diffusion noise are negligible in {\it E. coli} chemotaxis, orders of magnitude smaller than the internal noise floor (red symbols and lines in Fig. 3). Hence, these high frequency noises are unlikely the limiting factors for sensing accuracy in {\it E. coli} as usually assumed \cite{nwgprlramp,piglesias}. However, in other systems such as in eukaryotic chemotaxis, the binding time can be comparable to the response time $\tau_b\sim\tau_y$ \cite{bindeuk,cellstep}, so the ligand binding-unbinding noise $\eta_b$ can be significant. The ligand diffusion noise $\eta_d$ can also be more relevant for systems with relatively large receptors, which lead to a longer correlation time $\tau_d$ of ligand diffusion noise.


As we have shown above, the effect of the high frequency signal noise due to ligand binding and diffusion on the output fluctuation is negligible. The dominant source of signal noise is the low-frequency signal fluctuation that originates from the random motion of the cell in a ligand gradient. Here, we characterize this previously unstudied signal noise. 
In an exponential attractant gradient $[L]=[L]_0\e^{rx }$, a cell moves in a biased random walk with a constant drift velocity $v_d$ plus a random-walk velocity $\eta_v$ with correlation time $\tau_v$, variance $\sigma_v^2$ and the spectrum $\langle\eta_v^2(\omega)\rangle=2\omega_{v}\sigma_{v}^2/(\omega^2+\omega_{v}^2)$ \cite{bergrwb,liliexp,lilisim}. The random displacement (position) $\eta_x=\int \eta_v dt$ of the cell due to its random walk leads to a signal fluctuation $\eta_{rw}=r\eta_x$, which diverges as $\eta_{rw}\sim\omega^{-1}$ at low frequency $\omega\to0$. This strong signal fluctuation at low frequency is attenuated by the transfer function which goes as $\chi\sim i\omega$ at $\omega\to0$ due to adaptation's additional role as a low frequency controller. As a result, the random-walk induced output variance remains finite:
\begin{equation}
\sigma_{Y,rw}^2= [Ny(1-a)]^2(r\sigma_{v}\tau_{v})^2\times\frac{\tau_m}{\tau_{v}}.
\label{brw}
\end{equation}
The adaptive control of the random-walk induced signal noise is evident from the above expression. In Eq. \ref{brw}, $(r\sigma_{v}\tau_{v})^2$ is the signal variance during one random-walk step. The variance of the signal increases linearly with time $\sim t/\tau_{v}$ due to the random-walk nature of the cell motion. For $t\ge \tau_m$, the increase in signal variation is stopped by the adaptation process, and the variation of the output saturates to be proportional to $\tau_m/\tau_{v}$.

Direct simulations of Eq. \ref{interact} using the pathway interactions (Eqs. \ref{fa}-\ref{fy}) and random-walk induced signal noise confirm this analysis. As shown in in Fig. 2A\&B (blue symbols and lines), the effects of the random-walk induced noise are independent of $\tau_z$, but increase with $\tau_m$, exactly as predicted from Eq. (\ref{brw}). Depending on the ligand concentration gradient, the effects of this random-walk noise can be quite significant. For a exponential gradient $r=1\mathrm{mm^{-1}}$, the drift velocity is $v_d\approx 2\mathrm{\mu m/s}$ \cite{liliexp}, and the cell velocity fluctuation can be estimated from a pathway-based simulation \cite{lilisim} $\sigma_v\approx 9 \mathrm{\mu m/s}$ and $\tau_v\approx 0.5\mathrm{s}$. From Eq. ({\ref{brw}), the resulting relative output fluctuation is $\sigma_{Y,rw}/y\sim10^{-1}$, which is much larger than the ligand binding and diffusion noise (see Fig. 2A\&B), and comparable to the internal noise (see Fig. 3). The cell's rotational diffusion also contributes to this noise. However, this contribution \cite{piglesias,bialrot} is relatively small because the rotational diffusion time $\tau_{rd}\sim 10\mathrm{s}$ is much longer than $\tau_{v}\approx 0.5\mathrm{s}$. 

The internal noises have been analyzed quantitatively for {\it E. coli} chemotaxis to compare with the signal noise (Fig. 3). Both internal noise filtering processes, adaptation (node $m$) and output degradation (node $y$), contribute significantly to the total intrinsic noise. The adaptation process causes a larger output fluctuation $\sigma_{Y,m}/y\sim 10^{-1}$, because its timescale is much longer than the response time $\tau_m\gg\tau_y$. Note that a low-frequency noise of this magnitude is required to explain the observed 1/f noise in the switching dynamics \cite{grinstu,cluzelexp}. It was shown in \cite{cluzelexp} that increasing the amount of the methyltransferase CheR reduces the output noise. This observation was explained in \cite{emonet} by the possible ultra-sensitive dependence of kinase activity $a$ on CheR, which in our model translates to a change in the prefactor $C_m$ of the adaptation induced noise. However, our study reveals an additional mechanism: increasing CheR reduces the adaptation time $\tau_m$, and therefore the noise is averaged by $\tau_y$ more effectively as shown in Fig. 3B. The other main intrinsic noise source is the output node $y$, and its contribution to $\sigma_Y$ is smaller $\sigma_{Y,y}/y\sim10^{-2}$. Since the underlying molecular process is CheY-P de-phosphorylation, this intrinsic noise is a Poisson-type noise, its strength only depends on the number of molecules (CheY-P level). This analytical result, as seen in Fig. 3A, is consistent with the experimental observation that over-expressing CheZ and CheY together keeps the output variance constant \cite{sournoise}.

The topology of a biological network and the choices of the key biochemical constants determine its functions. For the chemotaxis network, different functions can be identified in different frequency regimes of the transfer function $\chi(\omega)$ as shown in Fig. 4A. At low frequencies $(\omega<\omega_m)$ random-walk noise is controlled and the gradient of the signal is computed (see \cite{yuhaiberg,shimitu}). At very high frequencies $(\omega>\omega_y)$ ligand diffusion and binding noise are averaged out. With $\omega_m\ll\omega_y$, there is also an intermediate region $\omega_y <\omega< \omega_m$, wherein the responses to external signal are kept fast and strong as shown in Fig. 4B. The {\it E. coli} chemotaxis network may have evolved to optimize these critical functions in different frequency regimes. The general tradeoffs of changing $\omega_m$ and $\omega_y$ become clear from our analysis. Reducing $\omega_m$ increases the sensitivity for gradient sensing since the CheY-P change in response to an exponential ligand gradient is $\propto\omega_m^{-1}$ \cite{yuhaiberg,shimitu}. But smaller $\omega_m$ reduces the range of gradient sensing (see Fig. 4A, and \cite{yuhaiberg,shimitu}), and reduces the low-frequency noise control ability which is crucial for suppressing the random-walk induced noise. Reducing $\omega_y$ enhances the cell's ability in filtering high frequency noise. On the other hand, increasing $\omega_y$ increases the response speed and response strength. As shown in Fig. 4B, the response (CheY-P level) to a step change in ligand concentration exhibits a peaked response $\Delta y_P$ at a time $\tau_P$. The dependence of the peak response $\Delta y_P$ and the peak time $\tau_P$ on the output degradation rate $\omega_y$ can be analytically derived, and are plotted in the inset of Fig. 4B. As one can see, increasing  $\omega_y$ reduces the peak time and increases the peak height. 

\section{Conclusions and Discussions}
\label{sec:1}

That noise can be a limiting factor for the correct behavior of a cell is an idea that dates back to Berg \& Purcell\cite{berg}, who first proposed the time-averaging mechanism to reduce the effect of ligand diffusion noise in receptors. Their work has recently been extended and made more rigorous by several groups \cite{bialek,levine,nwgprl}. However, the exact molecular nature of the time-averaging mechanism remained unclear. Here, by studying the dynamics of a typical adaptive signaling network, we show unambiguously that the time-averaging mechanism for the high frequency ligand binding and diffusion noise is the output decay process, not the receptor adaptation as previously suggested\cite{nwgprlramp,piglesias}. This result is common to all adaptive sensory systems, independent of the network topology.

For the particular case of {\it E. coli} chemotaxis, various aspects of noise in the signaling pathway have been studied in recent studies\cite{nwgprlramp,tenwoldenewprl}. We discuss some of these works in light of our findings in this paper. In \cite{tenwoldenewprl}, a signal-to-noise ratio study was carried out, with the ``signal" taken to be the response to a white noise \cite{tenwoldeoldprl}. This unrealistic assumption of the signal leads to the conclusion that information was best encoded at very high frequencies\cite{tenwoldenewprl}. However, it is well known that the chemotaxis pathway enables cells to detect deterministic changes (e.g., gradients) in chemo-effector concentration instead of random signals. As shown in this paper, the useful information resides at low (for ligand gradients) and intermediate (for ligand step change) frequencies (Fig. 4A). The noises caused by ligand binding and diffusion do have high frequencies, but they do not contain any useful information and are filtered out effectively by the Purcell-Berg time-averaging mechanism. In another recent work \cite{nwgprlramp}, a linear ligand gradient was considered as the input signal. However, this study claimed that the adaptation time is the noise-averaging time, probably because it considered the unrealistic case of having response time and adaptation time being the same. Moreover, although a linear ligand gradient was considered in \cite{nwgprlramp}, it did not identify the signal noise from the random-walk of the cell in a ligand gradient. As shown here in this paper, the randomness of in {\it E. coli} motion is in fact the main source of signal noise in a ligand gradient, while the signal noises caused by ligand diffusion and ligand binding considered before \cite{nwgprl,nwgprlramp,piglesias} are much smaller in comparison (Fig. 2). This dominant random-walk induced signal noise has a divergent low-frequency spectrum, which is fundamentally different from the simple white noise considered in previous studies \cite{nwgprlramp,tenwoldeoldprl}. We have shown that this previously un-characterized low-frequency noise is controlled by receptor adaptation.

For all adaptive networks, the low frequency noise is controlled by adaptation, while the high frequency noise is filtered by time-averaging. 
The characteristics of these two noise filtering strategies are fundamentally different. Time averaging works better for {\it longer} response times (Fig. 2A), but adaptive control works better for {\it shorter} adaptation times (Fig. 2B). However, the effectiveness in noise filtering, e.g., by {\it reducing} the adaptation time and {\it increasing} the response time, comes at the expense of the systems's ability to respond and sense. For instance, reducing the adaptation time gives a weaker response for gradient-sensing \cite{yuhaiberg}, and increasing the response time gives a weaker and slower response to a step change in input. These tradeoffs, partly identified in previous works \cite{piglesias}, are balanced in bacterial chemotaxis so that all the desired functionalities of the pathway come into play in different frequency domains as shown in Fig. 4A. At low frequencies $\omega<\omega_m$, adaptation allows gradient sensing and is also crucial in controlling the divergent random-walk induced signal noise. The speed and strength of step responses remains optimum for a range of intermediate frequencies ($\omega_m<\omega<\omega_y$). The high frequency ($\omega>\omega_y$) noise, such as the ligand binding and diffusion noise, are heavily suppressed by the Berg-Purcell time-averaging mechanism through output degradation.

\begin{acknowledgements}
This work was partially supported by a NIH grant (R01GM081747) to YT and a Cajamadrid fellowship to PS. We thank Dr. Ganhui Lan and Dr. Tom Shimizu for useful discussions, and Prof. Howard Berg for his careful reading of the first manuscript.
\end{acknowledgements}


\newpage

\begin{figure}[top]
\begin{center}
\includegraphics[width=\textwidth]{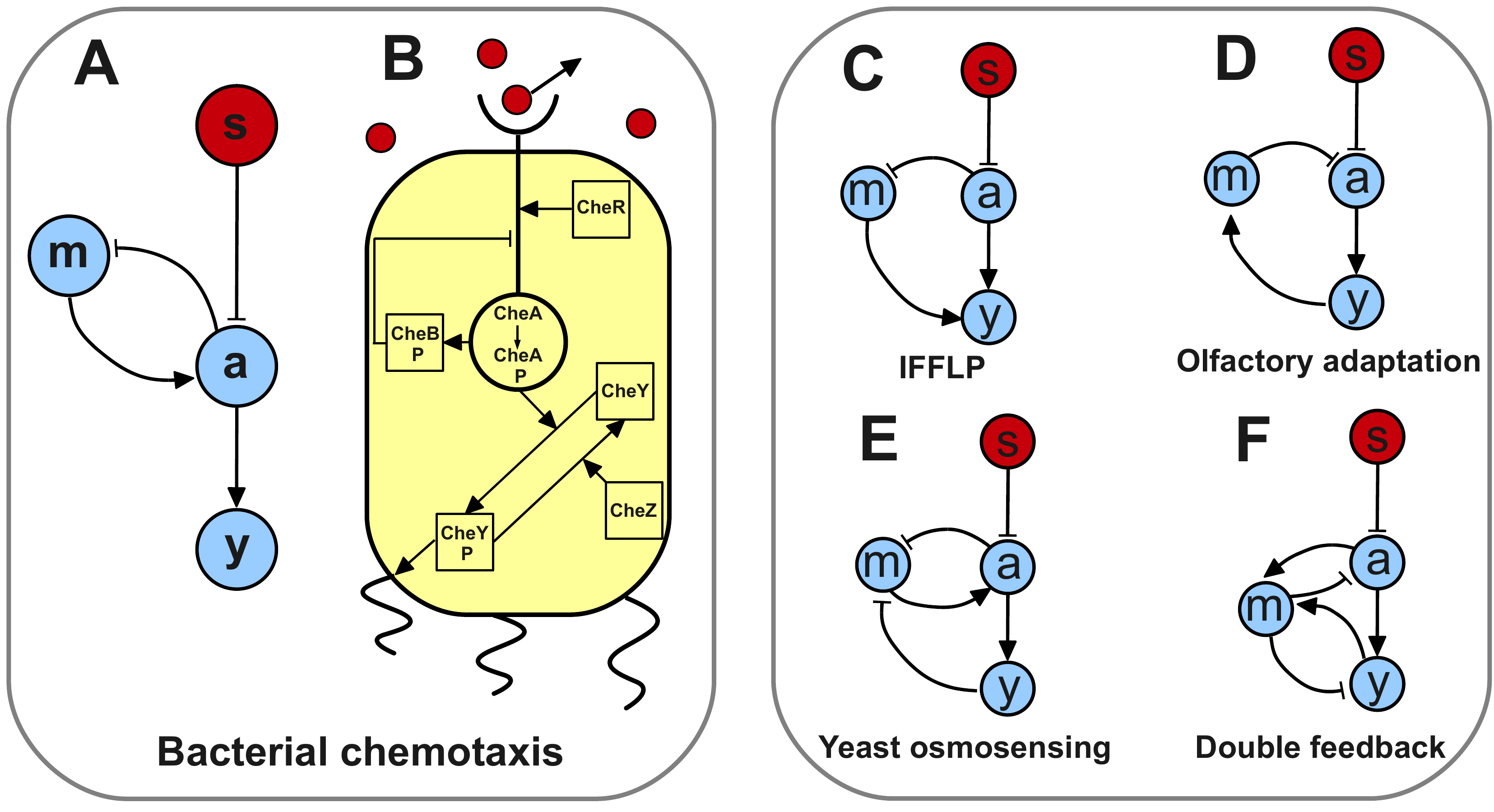}
\caption{Adaptive signaling networks capable of accurate adaptation: \textbf{A.} An illustration of a 3-nodes network capable of achieving perfect adaptation to step changes in the external signal $s$ (red circle). The signaling nodes (blue circles) are the readout $a$, the memory $m$ and the output node $y$. This is an abstraction of the \textit{E. coli} chemotaxis pathway. \textbf{B.} Signal transduction pathway for chemotaxis in \textit{E. coli}. \textbf{C-F.} Different topologies which can achieve accurate adaptation: \textbf{C.} Integrative Feed-Forward Loop with a Proportioner node (IFFLP) topology, identified in \cite{chaotang} as the main alternative network to the  chemotaxis topology in Fig. 1A, all other topologies are a composite of Fig. 1A and Fig. 1C. \textbf{D.} Topology of the olfactory sensing pathway in mammalian neurons. \textbf{E.} Topology identified in \cite{vanosmo} for yeast osmotic shock response. \textbf{F.} Adaptive topology that exhibits two negative feedbacks couple to the same memory.}
\end{center}
\end{figure}

\begin{figure}[top]
\begin{center}
\includegraphics[width=\textwidth]{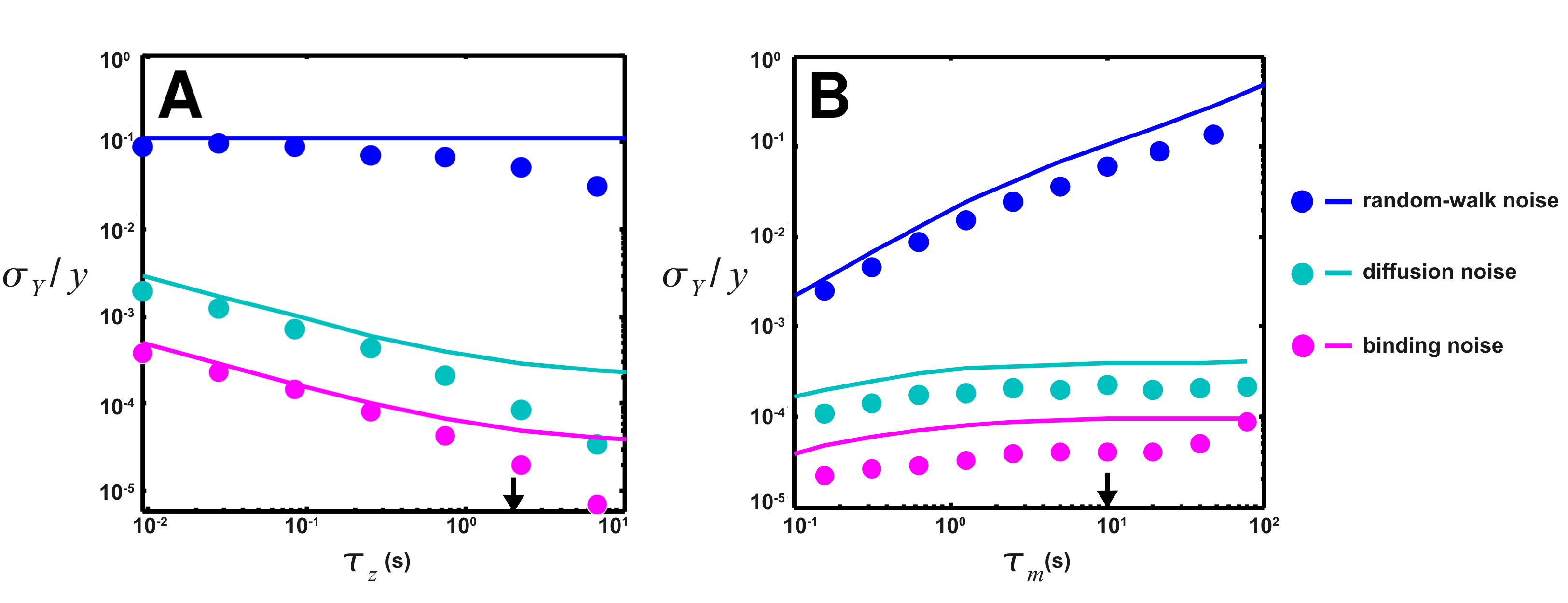}
\caption{Dependence of the three sources of signal noise on the output decay time ($\tau_z$) and the adaptation time ($\tau_m$). Results from direct simulations of Eq. \ref{interact} (symbols) and linear analysis (lines) are shown (see Eqs. \ref{fa}, \ref{fm} \& \ref{fy} for interactions and text for parameters used). Arrows indicate wild-type values of $\tau_z$ or $\tau_m$). The dependence of ligand binding noise (purple), ligand diffusion noise (light blue), and random-walk noise (dark blue) on \textbf{A.} $\tau_z$ and \textbf{B.} $\tau_m$. The ligand binding and diffusion noise drop much more than random walk noise when $\tau_z$ increases, while random walk noise drops much more than ligand diffusion noise when $\tau_m$ decreases. Because of their very short timescales, the effects of ligand binding and diffusion noise are much smaller than that of the random-walk noise, which is the dominant source of signal noise.
}
\end{center}
\end{figure}

\begin{figure}[top]
\begin{center}
\includegraphics[width=\textwidth]{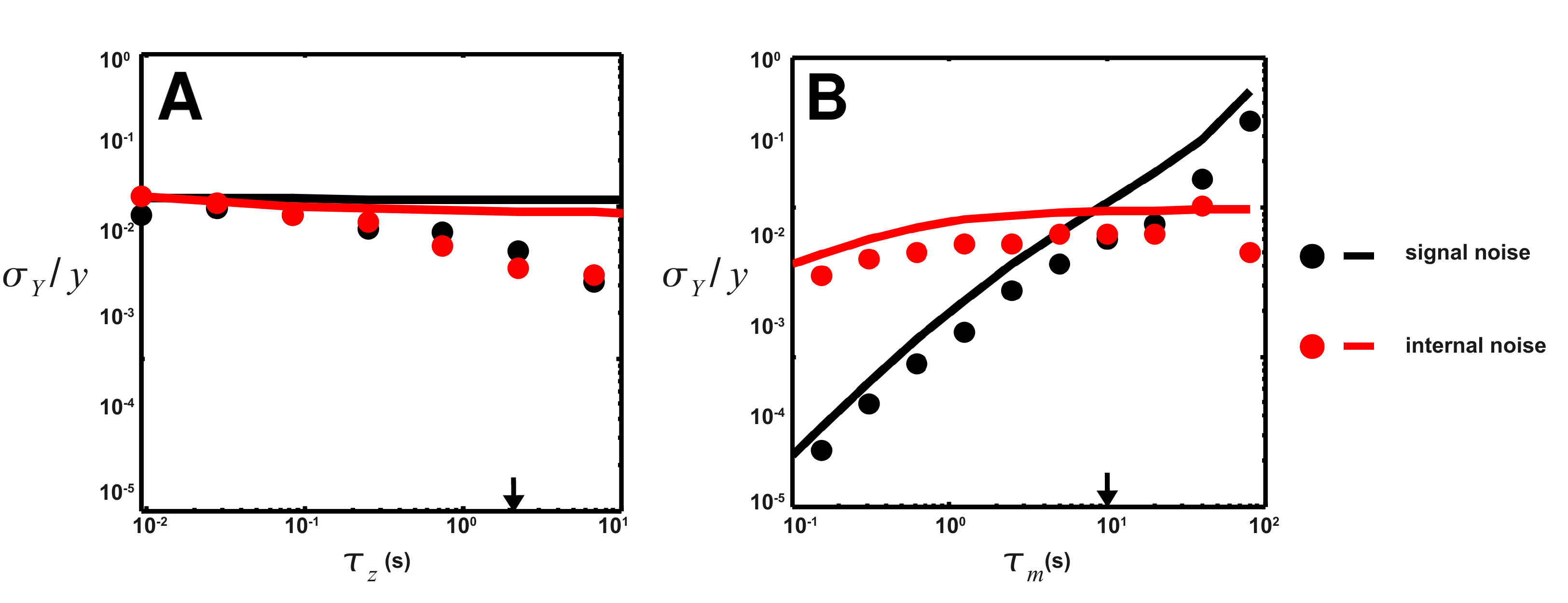}
\caption{Dependence of the total signal noise and the total internal noise on decay and adaptation times.  The dependence of the total internal noise floor (red) and the total signal noise (black) on \textbf{A.} $\tau_z$ and \textbf{B.} $\tau_m$. Signal noise is comparable to internal noise. At low $\tau_m\sim\tau_z$ adaptation noise starts being reduced by time averaging and the internal noise drops. Deviations between simulations (symbols) and linear analysis (lines) come from strong nonlinear effects when $\tau_z/\tau_y\gg1$. }
\end{center}
\end{figure}

\begin{figure}[top]
\begin{center}
\includegraphics[width=\textwidth]{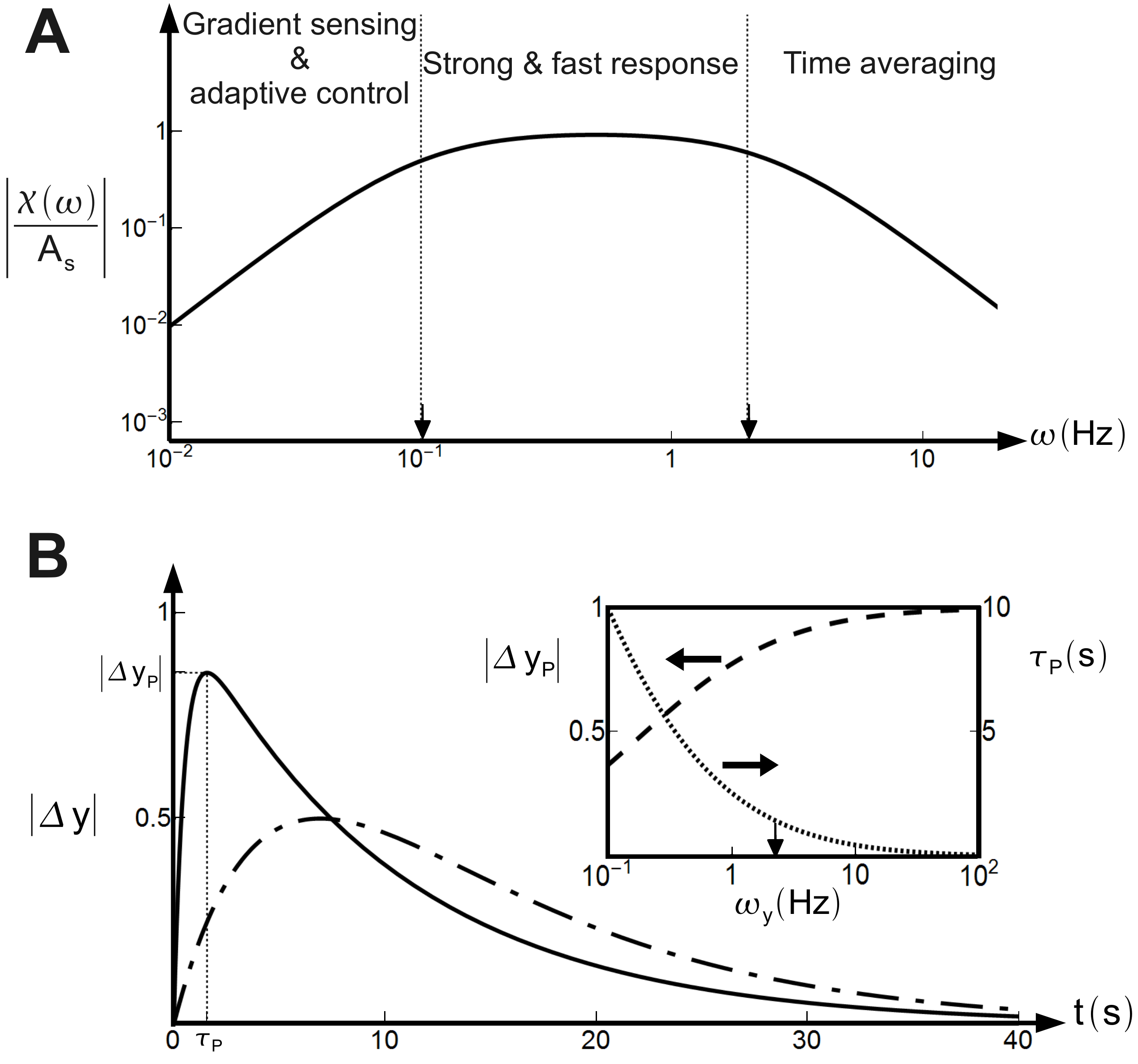}
\caption{The transfer function and response characteristics of adaptive networks. \textbf{A.} Three regimes can be identified in the transfer function (solid line) for \textit{E. coli} chemotaxis: gradient sensing and adaptive noise control is performed at low frequencies ($\omega<\omega_m$), response amplitude is strong for higher frequencies ($\omega_m<\omega<\omega_y$), and inputs with very high frequencies ($\omega>\omega_y$) are time averaged (axes arrows indicate wild-type values of $\omega_y$ and $\omega_m$). \textbf{B.} Output response $|\Delta y|$ to a small step change in the input. For fast response $\omega_{y}=20\omega_{m}$ as in \textit{E. coli} chemotaxis, the response is fast and strong (solid line). For slow response $\omega_{y}=2\omega_{m}$ as in eukaryotic chemotaxis, the response is slower and weaker (dot-dashed line). \textbf{Inset.} Dependence of the peaking time $\tau_P$ (dotted line) and peak height $|\Delta y_P|$ (dashed line) on $\omega_y$. For \textit{E. coli} wild-type value (axes arrow), both are near saturation.}
\end{center}
\end{figure}

\end{document}